%% file: main.tex
\documentclass[journal]{vgtc}                % final (journal style)
\ifpdf%                                % if we use pdflatex
  \pdfoutput=1\relax                   % create PDFs from pdfLaTeX
  \usepackage{graphicx}                % allow us to embed graphics files
  \DeclareGraphicsExtensions{.pdf,.png,.jpg,.jpeg} % for pdflatex we expect .pdf, .png, or .jpg files
\else%                                 % else we use pure latex
  \usepackage{graphicx}                % allow us to embed graphics files
  \DeclareGraphicsExtensions{.eps}     % for pure latex we expect eps files
\fi%

\usepackage{mathptmx}
\usepackage{graphicx}
\usepackage{times}
\usepackage{balance}

\usepackage{subfig}

\usepackage{amssymb}

\usepackage{multirow}

\usepackage{wrapfig}

\usepackage{enumitem}

\usepackage{booktabs}

\usepackage{bm}

\usepackage{tablefootnote}

\usepackage{xcolor}
\definecolor{lightgray}{rgb}{0.95, 0.95, 0.95}
\definecolor{mediumgray}{rgb}{0.75, 0.75, 0.75}
\definecolor{lightblue}{rgb}{0.95, 0.95, 1.00}

\definecolor{keywordcolor}{rgb}{0.25, 0.25, 0.25}
\definecolor{stringcolor}{rgb}{0.75, 0.25, 0.25}
\definecolor{commentcolor}{rgb}{0.25, 0.75, 0.25}

\usepackage[pdftex]{hyperref}
\definecolor{linkColor}{RGB}{6,125,233}
\hypersetup{
  pdfauthor = {},
  pdftitle = {},
  pdfsubject = {},
  pdfkeywords = {},
  colorlinks=true,
  linkcolor= black,
  citecolor= black,
  pageanchor=true,
  urlcolor=linkColor,
  plainpages = false,
  breaklinks=true,
}

\usepackage{ccicons}
\usepackage[colorinlistoftodos]{todonotes}
\usepackage{xspace}
\usepackage[normalem]{ulem}
\usepackage{enumitem}

\vgtcinsertpkg
\onlineid{XXX}
\vgtccategory{Research}

\usepackage{fancyhdr}
\definecolor{jungu}{rgb}   {0.1,0.3,0.8}
\definecolor{deok}{rgb}    {0.6,0.0,0.8}
\definecolor{niklas}{rgb}  {0.6,0.0,0.8}

\definecolor{amira}{rgb}   {0.0,0.3,0.67}
\definecolor{fanny}{rgb}   {0.4,0.1,0.7}
\definecolor{nicolas}{rgb} {0.8,0.6,0.0}
\definecolor{jacob}{rgb} {0.5,0.0,0.0}

\definecolor{fixme}{rgb} {1.0,0.0,0.0}

\newenvironment{SUBENVfixme}{\color{fixme}}{\color{black}}
\newenvironment{SUBENVcomment}[2]{\color{#1}[#2:~}{]\color{black}}

\definecolor{light-gray}{gray}{0.75}
\newcommand{\graytext}[1]{\textcolor{light-gray}{#1}}

\newcommand{\V}[2]{\textrm{V}_{#1}(#2)}
\newcommand{\Size}[2]{\textrm{S}_{#1}(#2)}
\newcommand{\Pos}[2]{\textrm{P}_{#1}(#2)}
\newcommand{\Lum}[2]{\textrm{L}_{#1}(#2)}

\renewcommand{\quote}[1]{``\textit{#1}''}

\newcommand\eg{e.g.\xspace}

\renewcommand\SS{\xspace{\sc ss}\xspace}
\newcommand\DS{\xspace{\sc ds}\xspace}
\newcommand\DP{\xspace{\sc dp}\xspace}
\newcommand\DL{\xspace{\sc dl}\xspace}
\newcommand\SP{\xspace{\sc sp}\xspace}
\newcommand\SL{\xspace{\sc sl}\xspace}

\newcommand\surveyURL{\url{http://bit.ly/common-fate}}

\title{Common Fate for Animated Transitions in Visualization}

\author{Amira Chalbi,$^\dagger$ Jacob Ritchie,$^\dagger$ Deokgun Park, Jungu Choi,\\ Nicolas Roussel, Niklas Elmqvist, \textit{Senior Member, IEEE}, and Fanny Chevalier}
\authorfooter{

\item[$\dagger$] The first two authors contributed equally to the work.

\item Amira Chalbi and Nicolas Roussel are with Inria, France.  E-mail: amira.chalbi.ihm@gmail.com, nicolas.roussel@inria.fr.

\item Jacob Ritchie and Fanny Chevalier are with the University of Toronto in Toronto, ON, Canada.  E-mail: jritchie@dgp.toronto.edu, fanny@cs.toronto.edu.

\item Deokgun Park is with the University of Texas at Arlington, TX, USA.  E-mail: deokgun.park@uta.edu.

\item Jungu Choi is with Purdue University in West Lafayette, IN, USA.  E-mail: monsieur.choix@gmail.com.

\item Niklas Elmqvist is with the University of Maryland in College Park, MD, USA.  E-mail: elm@umd.edu.
}

\abstract{The Law of Common Fate from Gestalt psychology states that visual objects moving with the same velocity along parallel trajectories will be perceived by a human observer as grouped.
However, the concept of \emph{common fate} is much broader than mere velocity; in this paper we explore how common fate results from coordinated changes in luminance and size.
We present results from a crowdsourced graphical perception study where we asked workers to make perceptual judgments on a series of trials involving four graphical objects under the influence of conflicting static and dynamic visual factors (position, size and luminance) used in conjunction.
Our results yield the following rankings for visual grouping: motion $>$ (dynamic luminance, size, luminance); dynamic~size~$>$~(dynamic luminance, position); and dynamic luminance~$>$ size.
We also conducted a follow-up experiment to evaluate the three dynamic visual factors in a more ecologically valid setting, using both a Gapminder-like animated scatterplot and a thematic map of election data.
The results indicate that in practice the relative grouping strengths of these factors may depend on various parameters including the visualization characteristics and the underlying data.
We discuss design implications for animated transitions in data visualization.}


\begin{document}

\maketitle

%% --------------------------------------------------------------------- 
%% Introduction
%% ---------------------------------------------------------------------
\input{01-intro}

%% ---------------------------------------------------------------------
%% Related Work
%% ---------------------------------------------------------------------
\input{02-background}

%% ---------------------------------------------------------------------
%% Common Fate
%% ---------------------------------------------------------------------
\input{03-common-fate}

%% ---------------------------------------------------------------------
%% User Study Background
%% ---------------------------------------------------------------------
\input{04-user-study-background}

%% ---------------------------------------------------------------------
%% User Study
%% ---------------------------------------------------------------------
\input{05-user-study}

%% ---------------------------------------------------------------------
%% Results
%% ---------------------------------------------------------------------
\input{06-results}

%% ---------------------------------------------------------------------
%% Followup study
%% ---------------------------------------------------------------------
\input{07-followup}

%% ---------------------------------------------------------------------
%% Discussion and Implications 
%% ---------------------------------------------------------------------
\input{08-disc}

%% ---------------------------------------------------------------------
%% Conclusion & Future Work
%% ---------------------------------------------------------------------
\input{09-conclusion}

%% ---------------------------------------------------------------------
%% Acknowledgments
%% ---------------------------------------------------------------------
\section*{Acknowledgments}

We thank Pierre Dragicevic and Dan Elliott for useful discussions on this project, Alex Quinn, Alix Goguey, and Bruno De Araujo for comments on the experiment, and Kai Cheng Thom for the typing.
We also thank the crowd workers for their labour and the anonymous reviewers for their constructive feedback. 
Parts of this work previously appeared in Amira Chalbi's doctoral thesis~\cite{Amira_Diss} (2018).

%\newpage
%% ---------------------------------------------------------------------
%% References
%% ---------------------------------------------------------------------
\balance{}
\bibliographystyle{abbrv-doi}
\bibliography{main}

\end{document}

%% file: 01-intro.tex
%% --------------------------------------------------------------------- 
%% Introduction
%% ---------------------------------------------------------------------
\section{Introduction}

Animation is commonly used for state changes in HCI and visualization applications, allowing the viewer to gradually track changes in an interface rather than having to reinterpret a visual representation or interface from scratch~\cite{Baecker1990, Robertson1989}.
However, designing animated transitions so that they convey changes that are smooth and simple to follow is not trivial, and involves issues such as pacing~\cite{Dragicevic2011}, staging~\cite{Chevalier2014}, and tracking~\cite{Pylyshyn1988, Tversky2002} of animated objects.
The Gestalt Law of Common Fate (LCF)~\cite{Koffka1922} is an example of a widely known guideline for designing animations, where visual elements that move with the same velocity (i.e.\ same speed and same direction) are said to be perceived as sharing the same ``fate'', and thus belong to the same group.
The LCF is also the only of the five Gestalt Laws that deals with dynamic (i.e.\ animated, time-changing) properties; the others all concern static instances of grouping in visual perception~\cite{Wagemans2012}.

Although the Gestalt Laws---including LCF---were derived from perceptual psychology experiments in the early 1900s at the ``Berlin School'' of psychology~\cite{Koffka1922}, only a few isolated examples of application to \emph{dynamic} visualizations have been explored~\cite{Blok1999, Friedrich2002, Ware2005}.
This presents an opportunity for visualization research to delve deeper into human perception for the purposes of optimizing animated transitions.
For example, better knowledge of the automatic grouping of animated objects may suggest ways to structure animated transitions so that their complexity is decreased and they become easier to perceive.
Furthermore, while most examples of LCF use visual elements with identical trajectories~\cite{Koffka1922}, the philosophical meaning of a ``common fate'' of objects engaged in joint motion is not necessarily restricted to velocity~\cite{Sekuler2001}.
Rather, a general interpretation of ``common fate'' might merely imply shared dynamic behavior between multiple objects that creates a perception that they are under the influence of the same physical process~\cite{Alais1998}.
Such shared behaviors include growth and compression (size) as well as darkening and brightening (luminance).
Given this background, it is useful to ask ourselves how the visual grouping arising from common fate is influenced by such dynamic behaviors, and how these factors interact with each other.
Answering these questions may shed light on potential new ways to add structure to animated transitions in interfaces and visualizations.

In this paper, we study these intricacies of the Gestalt Law of Common Fate by means of a large-scale online graphical perception experiment involving 100 crowdworkers performing perceptual grouping tasks.
Our experiment was designed to compare three static visual factors (position, size, and luminance) and three dynamic visual factors (velocity, luminance change, and size change).
For each trial, four graphical objects were grouped by two properties at a time: two pairs were grouped based on one factor and two other pairs based on another.
This enabled us to not only study the individual grouping strength of each visual factor, but also to rank the factors in order of their relative grouping strength.
Furthermore, to increase the ecological validity of our work, we also conducted a follow-up experiment asking participants to perceive dynamic changes in an animated scatterplot as well as a thematic map.
We discuss how these findings can inform the design of animated transitions to reduce cognitive load.

%% file: 02-background.tex
%% --------------------------------------------------------------------- 
%% Related Work
%% ---------------------------------------------------------------------
\section{Background}

Here we give a general overview of relevant work in perceptual psychology, graphical perception, and animation in visualization.

\subsection{Perception and Gestalt Psychology}

%\jacob{CHI 2015 reviews have comments that indicate that they disagree with some of the statements in this section.}
\textit{Perception} comprises the innate sensory components of the human cognitive system that are pre-conscious and used to represent and understand the environment, and \textit{visual perception} is the perceptual component dealing with sight.
As the most important of the senses, the human visual system has evolved over millions of years to allow individuals to distinguish, identify, and track objects in their vision~\cite{Jerison1973}.
%Unlike the brain, which even with today's technology is difficult to observe and decipher, the sensory organs are more straightforward, which led to much of early cognitive psychology experiments focusing on perception.
%Visual perception is no exception.

Much of the seminal work on visual perception was conducted in the early 1900s by the so-called ``Berlin School'' of experimental psychology.
This eventually led to the development of \textit{Gestalt psychology}~\cite{Koffka1922}, a theory of mind based on a holistic view of human visual perception where the sum of the perceived ``gestalt'' is qualitatively different than its component parts, and in effect has an identity of its own.
One key practical outcome of Gestalt psychology was the development of the law of \textit{pr{\"a}gnanz} (German, \textit{pithiness})~\cite{Koffka1922}, which can be operationalized into the so-called ``Gestalt laws''~\cite{Wagemans2012}: examples include the Law of Proximity, which states that objects at close distances are perceptually grouped, or the Law of Similarity, which states that objects with similar visual appearance are grouped together.
Analogously, the Law of Common Fate---incidentally, the only Gestalt law dealing with dynamic settings--- is commonly understood to state that objects with the same movement are perceptually grouped together. Recent research suggests that the same feature selection mechanism may underlie both similarity and LCF-based grouping \cite{Levinthal2011, Yu2019}. 
% \jacob{Move references to generalized common fate studies in vision science up here}

\subsection{Motion and Animation}

Much of visual perception evolved for survival purposes, and few perceptual properties have the urgency of rapid movement.
As a result, the human visual system is highly sensitive to motion and is capable of tracking multiple objects moving simultaneously~\cite{Cavanagh2005, Pylyshyn1988}.
Animation, where the illusion of motion is recreated through the rapid display of a sequence of static images, is thus of interest  both for psychology and for entertainment applications in artificial settings.

Animation has long been used in graphical user interfaces to show progress, convey state transitions, and notify the user of changes~\cite{Baecker1990, Hudson1993,DBLP:conf/avi/ChevalierRPCH16}.
Animated transitions have become particularly popular and are used in a variety of applications, ranging from presentation software and video editors to visualization tools~\cite{Fisher2010}.
Perceptual studies suggest that smooth transitions not only improve user decision-making~\cite{Gonzalez1996}, but also facilitate their mental map~\cite{Bederson1999} and recall~\cite{Shanmugasundaram2008}.
Despite much literature praising the merits of animations, Tversky et al.~\cite{Tversky2002} note that there exist several studies showing that they could harm more than they help, but attribute the unpromising conclusions to poor animation design choices and flaws in evaluation protocols.

Animations can be specifically designed to convey data.
Cartoonists were the first to investigate how to communicate emotions through motion~\cite{Johnston1995}. 
Bartram et al.~\cite{Bartram2003} proved the effectiveness of animated icons to notify changes.
Ware et al.~\cite{Ware1999} suggested the use of animation to express causal relationships between entities in visualizations.

Structured animations can be used to reduce visual complexity during a transition. 
Heer and Robertson~\cite{Heer2007} proposed introducing discrete stages during transitions between statistical data graphics to help users follow the animations.
Chevalier et al.~\cite{Chevalier2014} investigated \textit{staggering}---an extreme case of staging---but found no positive impact on user performance.
Dragicevic et al.~\cite{Dragicevic2011} studied how temporal pacing for animation can be distorted to improve perception, and Du et al.~\cite{Du2015} studied how a spatio-temporal structuring can reduce visual complexity by bundling  trajectories of animated objects.

Animated transitions are also increasingly used in information visualization to support various operations such as filtering, sorting, zooming, or changing visual representations.
Bartram and Ware were among the first to study them in this context for brushing~\cite{Bartram2002}.
Van Wijk and Nuij~\cite{VanWijk2004} proposed a mathematically optimized animation scheme for panning and zooming so that the visual flow is invariant.
The ScatterDice~\cite{Elmqvist2008} and GraphDice~\cite{Bezerianos2010} techniques leverage shape transitions between scatterplots and node-link diagrams, respectively.

\subsection{Gestalt Laws in Visualization}

The Gestalt Laws are an important component of visual perception that researchers have attempted to leverage for more efficient visual communication.
Early work in cartographic animation assumes that common fate is more generally valid for objects that change together, e.g., by blinking, though no formal evaluation is reported~\cite{Blok1999}.
Ware and Bobrow used motion as a mechanism to highlight a subgraph of interest in a larger graph.
While they initially found that motion dominates hue for highlighting~\cite{Ware2004}, their most recent studies suggest that motion and hue can be used in conjunction for the highlighting of two different entities simultaneously~\cite{Ware2005}.
Finally, Romat et al.~\cite{Romat2018} can be said to leverage the notion of common fate by animating the link lines in a node-link diagrams to indicate direction, rate, and speed.

Friedrich and colleagues~\cite{Friedrich2002, Nesbitt2002} successfully applied the Law of Common Fate to make subgraphs apparent when transitioning from one layout to another to preserve the viewer's mental map.
The goal was to find an animation of the subgraph of interest that would be interpreted by the brain as movement of three-dimensional objects, using affine transforms to decompose the motion into a series of translations, rotations, scalings, and shears.
They found that the Law of Common Fate not only holds for objects moving in the same direction, but also for objects which move in any structured way.
However, none of these studies are empirical, and, more importantly, few prior efforts have directly studied common fate for visualization.

%% file: 03-common-fate.tex
%% --------------------------------------------------------------------- 
%% Common Fate
%% ---------------------------------------------------------------------
\section{The Gestalt Law of Common Fate}
\label{sec:common-fate}

The predominant interpretation of the Gestalt Law of Common Fate (LCF) is that the concept of ``common fate'' solely refers to the visual grouping of elements moving in a \emph{coherent motion}, i.e.\ with the same speed and direction.
One way to intuitively explain this phenomenon is that the moving objects that are visually grouped are under the influence of a single factor causing them to move along the same trajectory.

However, this simplistic interpretation is not the only one.
Wertheimer, one of the founders of Gestalt psychology, used moving objects with identical velocity as an illustrating example in his original German manuscript~\cite{Wertheimer1923}.
However, as noted by Sekuler and Bennett in 2001~\cite{Sekuler2001}, he also included a passage on broader interpretations of the concept of common fate that never appeared in the English transcript: \quote{The principle [of common fate] applies to a wide range of conditions; how wide, is not discussed here.}

Biased by the belief that Gestaltists only had motion in mind when developing LCF, subsequent studies in psychology have mostly focused on investigating the limits of figure-ground segmentation under variations of motion coherence~\cite{Lee2001, Sturzel2004, Uttal2000}, which may explain why the simplified and incomplete version of the law has become prevalent.
Exceptions include studies on dynamic luminance~\cite{Alais1998, Sekuler2001} and its informal application to cartographic animation~\cite{Blok1999} and graph visualization~\cite{Nesbitt2002, Ware2005}.
As stated by Brooks in a recent survey on perceptual grouping: \quote{Although common fate grouping is often considered to be very strong, to my knowledge, there are no quantitative comparisons of its strength with other grouping principles.}(p.~60,~\cite{Brooks2014})

Given this background, we formulate two distinct research questions that we focus on in this work:

\begin{itemize}[itemsep=0pt, topsep=0pt]

\item[RQ1]\textit{Does the Law of Common Fate extend to other dynamic visual variables, such as dynamic luminance or size?}
  While past work~\cite{Alais1998, Sekuler2001} has proved this for luminance, we want to study this more broadly for other visual variables.
  
\item[RQ2]\textit{What is the relation between the (extended) Law of Common Fate and other Gestalt Laws?} 
  As the only Gestalt law dealing with animation, and given the perceptual urgency of motion~\cite{Tversky2002}, we are interested in the relation between LCF and other Gestalt laws.
  
\end{itemize}

To answer these, we discuss criteria that may have an impact on perceptual grouping and identify visual variables that obey them.

\subsection{Criteria for Perceptual Grouping}

As is clear from the above treatment of Gestalt psychology, perceptual grouping of visual objects arises from relations between \textit{visual variables}.
Identifying (and ranking) such visual variables was one of the fundamental advances of early work in visualization; for example, Bertin~\cite{Bertin1967} lists seven visual variables, and Cleveland and McGill~\cite{Cleveland1985} list ten.
However, it is not feasible for us to study all of these visual variables, and besides, not all of them have the same potential for exhibiting perceptual grouping.
Here we describe our selection criteria.

\subsubsection{Associativity}

Visual variables that support grouping are often described as \emph{associative} in the literature. 
It is worth noting, however, that this term has often been misunderstood by the community.
As Carpendale points out~\cite{Carpendale2003}, there are discrepancies between the notion of associativity as defined by Bertin~\cite[p.\ 48]{Bertin1967} and that which is usually understood~\cite{Reimer2011}.
For Bertin, a variable is associative if objects can be grouped across other variables despite changes in that one.
In contrast, Carpendale's definition of associativity refers to perceptual grouping power.

Since we focus on grouping, we adopted Carpendale's definition, yielding one dynamic (motion) and eight static (position, size, shape, luminance, color, orientation, grain, texture) associative variables.

\subsubsection{Ordered Transitions}

Since our focus is on common fate, our second criterion of selection pertains to the dynamic aspects of the above listed associative variables.
Among these variables, there are several for which it is difficult to describe a dynamic behavior and specify a transition.
For instance, working with shapes, textures or color hue, we would have many options to choose from as people (including us) have no clear intuition of how one should transition from one value to another.

Thus, in the spirit of keeping the study as simple as possible, we focused on variables that, in addition to being associative, are \emph{ordered}, i.e.\ a change in these variable can be perceptually interpreted as increasing or decreasing.
This allows for deterministically interpolating between values for both increasing transitions (the value of the visual variable grows), or decreasing transitions (the value becomes smaller).
From the list of associative variables given by Carpendale~\cite{Carpendale2003}, only position, luminance, and size are ordered.

\subsection{Visual Variables with Grouping}

We focus on the static and dynamic versions of the three visual variables satisfying our criteria as follows:

\subsubsection{Static Variables}

Static visual variables are invariant over time and thus do not create the perception of common fate.
However, including these factors allows us to answer RQ2 on the relation between LCF and other laws.

\begin{itemize}[itemsep=0pt, topsep=0pt]

\item\textit{Static position (\SP):} Visual elements in close proximity are perceived as grouped, a phenomenon known as the Law of Proximity.
Geometric position is also generally ranked as the most perceptually accurate visual variable~\cite{Bertin1967, Cleveland1985}.

\item\textit{Static size (\SS):} According to the Law of Similarity, elements with the same size will be grouped together. Bertin~\cite{Bertin1967} names size as the second most perceptually accurate visual variable, whereas Cleveland and McGill~\cite{Cleveland1985} rank area as number five.

\item\textit{Static luminance (\SL):} By the same Law of Similarity, elements with the same color are grouped.
Bertin ranks it at number five, and Cleveland and McGill rank it as ``color saturation'' at six.

\end{itemize}

\subsubsection{Dynamic Variables}

Dynamic visual properties represent behavior that changes over time, which means that they may exhibit common fate effects.
These factors allow us to answer RQ1 on whether the concept of common fate extends beyond mere object motion.

\begin{itemize}[itemsep=0pt, topsep=0pt]

\item\textit{Dynamic position (\DP):} The canonical example of the Law of Common Fate: objects moving with the same speed and direction are perceived as belonging to the same group.
    
\item\textit{Dynamic size (\DS):} Are visual objects that grow or shrink in the same manner perceived as belonging to the same group?

\item\textit{Dynamic luminance (\DL):} As shown by prior studies, visual objects becoming brighter or darker in the same way are perceived as belonging to the same group~\cite{Sekuler2001, Wagemans2012, Ware2005}.
However, these experiments did not allow for investigation of the relation of \DL to other visual variables, both static and dynamic.
  
\end{itemize}

%% file: 04-user-study-background.tex
%% ---------------------------------------------------------------------
%% Study Rationale
%% ---------------------------------------------------------------------
\section{Study Rationale}

Our goals are to \textit{(i)} determine whether the LCF extends to visual variables beyond motion, and to \textit{(ii)} determine the relative grouping strength of LCF and other Gestalt laws.
Here we present our rationale.

\subsection{Task Rationale}
\label{sec:rationale}

In our study, we chose to give participants perceptual tasks where four graphical objects were grouped by two properties at a time so as to create two orthogonal possible groupings, and ask participants which emergent groups they perceive.
In other words, we make two visual variables compete, and record which one---if any---coincides with the participant's answer, and hence influenced their grouping perception.

From a visual variable's grouping power perspective, any answer to the above question falls into one of the three following categories:
\emph{(i)} the participant's grouping coincides with that dictated by the first visual property, and we can assume that the corresponding visual variable thus has the highest grouping strength for this task;
\emph{(ii)} they grouped the objects based on the second, competing property, so we assume that the other visual variable has the highest grouping strength for this task; or \emph{(iii)} none of the above (i.e., they grouped differently), in which case none of the two variables can be said to have a grouping power for this task.

Our focus being on common fate, we are primarily interested in tasks where dynamic variables are involved, and hence on animated transitions implementing these dynamic behaviors.
However, for the sake of experimental completeness, we also tested static variables against each other, and our trials also included static visualizations.

By making a dynamic visual variable compete against any of the static variables whose grouping power is established (i.e., by the Law of Proximity or the Law of Similarity), we can quantitatively measure the grouping power of the Law of Common Fate---in our case, restricted to motion, dynamic luminance, or dynamic size.
The more cases where participants deviate from the Laws of Proximity and Similarity in favor of the dynamic property, the stronger the evidence that the associated dynamic visual variable has perceptual grouping power, and subsequently the stronger the evidence that the Law of Common Fate applies to this variable (RQ1).
The relative grouping strength between each variable is directly measurable from tasks comparing pairs of non-conflicting visual variables (RQ2).

\subsection{Summary of Tasks}
\label{sec:tasks}

Table~\ref{tab:tasks} summarizes all of the possible pairwise comparisons for our six visual variables.
Out of the 36 cells, we do not consider self-comparisons (diagonal), nor do we count duplicates (i.e.\ \SP~vs.~\DL is the same as \DL~vs.~\SP); these are grayed out.
We also discard any pairwise comparison where a dynamic visual variable competes against its static counterpart (e.g.\ \SS~vs.~\DS).
The reason is to avoid conflicts: having orthogonal groups bound to the same visual variable would necessarily break the notion of similarity at a point during the animation, making such cases difficult to interpret.\footnote{For example, comparing \SP\ and \DP\ would mean that two objects grouped by static position would only be in proximity during a single point in the trial, \eg at the beginning or end; they would become separated (and thus no longer near) by varied dynamic positions (velocities) during the rest of the trial.}

This leaves 12 distinct pairwise comparisons that form our set of tasks for the study: \DP-\DS, \DP-\DL, \DP-\SS, \DP-\SL, \DS-\DL, \DS-\SP, \DS-\SL, \DL-\SP, \DL-\SS, \SP-\SS, \SP-\SL,~and \SS-\SL.

\begin{table}[ht]
  \vspace{-3mm}
  \caption{Comparison Tasks Generated from the Six Visual Variables.}
  \centering
  \label{tab:tasks}
    \begin{tabular}{  l  c  c  c  c  c  c }

      & {\bf \DP} & {\bf \DS} & {\bf \DL} & {\bf \SP} & {\bf \SS} &
      {\bf \SL }\\
      
      {\bf \DP} & --- & \graytext{\DS-\DP} & \graytext{\DP-\DL} &
      \graytext{---} & \graytext{\DP-\SS} & \graytext{\DP-\SL} \\
      
      {\bf \DS} & \DP-\DS & --- & \graytext{\DS-\DL} &
      \graytext{\DS-\SP} & \graytext{---} & \graytext{\DS-\SL} \\
      
      {\bf \DL} & \DP-\DL & \DS-\DL & --- & \graytext{\DL-\SP} &
      \graytext{\DL-\SS} & \graytext{---} \\
      
      {\bf \SP} & --- & \DS-\SP & \DL-\SP & --- & \graytext{\SP-\SS} &
      \graytext{\SP-\SL} \\
      
      {\bf \SS} & \DP-\SS & --- & \DL-\SS & \SP-\SS & --- &
      \graytext{\SS-\SL} \\
      
      {\bf \SL} & \DP-\SL & \DS-\SL & --- & \SP-\SL & \SS-\SL & --- \\
    \end{tabular}
\end{table}
\vspace{-4mm}

\subsection{Manipulation of Visual Variables}

Let \textit{visual property} henceforth denote a specific value for a visual variable.
To create the above tasks, where objects are grouped by similar visual properties, we manipulate static properties (i.e.\ position, size, and luminance) as well as dynamic behaviors (i.e.\ changes in position, changes in size, and changes in luminance).
Through these manipulations across objects, we can manipulate the relation of \emph{similarity}---in the most general sense of the term---between objects to create distinct groups of objects sharing similar visual properties.

Here, we propose a generalization of \emph{similarity} in a particular visual variable's definition space for both the static and dynamic aspects.

\subsubsection{General Notation}

In the following, we use $\mathcal{S}$ to refer to the set of visual objects in a task.
For a given object $A$ in $\mathcal{S}$, $\V{A}{t}$ refers to the value of a visual variable at time $t$, and $\Delta\V{A}{t_{i-1},t_{i}}$ denotes the difference of values for $A$ between time $t_{i-1}$ and $t_{i}$ (i.e.\  $\Delta\V{A}{t_{i-1},t_{i}}~=~\V{A}{t_{i}}~-~\V{A}{t_{i-1}}$), where the increment between times $t_{i-1}$ and $t_{i}$ corresponds to one step at the finest observable temporal resolution.

Let $\Pos{A}{t}$, $\Size{A}{t}$ and $\Lum{A}{t}$ refer to the position, size, and luminance of the object $A$ at time $t$ of the animation. Object luminance is normalized to $[0, 1]$, where 0 is black and 1 is white, for a given display.

\subsubsection{Similarity and Similar Behavior}
\label{sec:similarity}

Visual objects $A$ and $B$ are \textbf{similar} with respect to a visual variable $\mathrm{V}$ at time $t$ if their difference is below a threshold:
%\vspace*{-0.5em}
%\begin{center}
$ |\V{A}{t} - \V{B}{t}| \leq \tau_{\textrm{V}}$
%\end{center}
%\vspace*{-0.5em}  
  
In the static case, the notion of similarity for two objects $A$ and $B$ directly refers to the Law of Proximity for position, and the Law of Similarity for size and luminance.
In other words, these static situations correspond to the special cases in the above definition where $\Pos{A}{t}$ and $\Pos{B}{t}$, $\Size{A}{t}$ and $\Size{B}{t}$, $\Lum{A}{t}$ and $\Lum{B}{t}$ are constant over time (i.e., static position, size, and luminance). 

What the Law of Common Fate suggests, is that even if objects are not similar at any time $t$, the fact that they behave similarly is a factor for perceptual grouping.
Put differently, this means that the difference in their \emph{variations} across time is below a certain threshold.
Formally, visual objects $A$ and $B$ \textbf{behave similarly} between $t_{i-1}$ and $t_{i}$ if:

\vspace*{-0.5em}
\begin{center}
$|\Delta\V{A}{t_{i-1},t_i} -\ \Delta\V{B}{t_{i-1},t_i}| \leq \theta_{\textrm{V}}$
\end{center}
\vspace*{-0.5em}

Applying the above definitions in the context of our visual variables during an animated transition, we have:

\vspace*{-0.4em}
\begin{itemize}[noitemsep, label={}]
\item $A$ and $B$ are \SP-similar (resp. \SS-similar; \SL-similar) if: $A$ and $B$ are \textit{similar} in position (size; luminance), $\forall{t}$ of transition;
\vspace*{3pt}
\item $A$ and $B$ are \DP-similar (resp. \DS-similar; \DL-similar) if: $A$ and $B$ \textit{behave similarly} in position (size; luminance), $\forall{t}$ of transition.
\end{itemize}
\vspace*{-0.4em}

We can operationalize these rules to create groupings for any of the above visual variables by ensuring both that (1) objects that are to be grouped are indeed similar (within some tolerance), and that (2) there exist no other object in $\mathcal{S}$ that is similar to the objects in the group.
We note that these rules do not apply generally across all situations, but only in the context of our controlled experiment; in general, similarity is highly contextual.
For example, two objects with identical luminance will not be perceived as similar if one is placed on a darker background and the other on a lighter background.

\subsubsection{Neutrality}

To control for perceptual processes and confounding effects, all objects in $\mathcal{S}$ should be theoretically \emph{neutral}, i.e.\ they should all be similar in all aspects (both static and dynamic).
For simplicity and to guarantee perceptual grouping neutrality, we use a set of \emph{static} and \emph{identical} visual objects as a default set.
It is only when testing the effect of visual variables on grouping that we modify these specific object properties to create distinct groups, as described above.

The only exception for neutrality is position, since it does not make
sense to have objects overlap.
Perfect position similarity (i.e.\ $\tau_P=0$) would entail all objects sharing the exact same position.
In fact, we also cannot enforce equidistance between all possible pairs of objects for sets of more than three objects.
Dot lattices are commonly used in psychology experiments that study proximity grouping~\cite{Brooks2014}; however, we chose to avoid too much regularity in object arrangement since this can also lead to grouping by proximity~\cite{Strother2006}.

Any positioning strategy deviating from the above rules will necessarily introduce a small bias for a set of more than three objects.
To minimize the spatial proximity that may occur by uniform random positioning, we used a similar approach to Poisson-disc sampling~\cite{Bridson2007}, which results in a balanced spatial distribution by adding a constraint on the spatial position of each object relative to the closest neighbor: each object must be located within a distance range $[d_{min}, d_{max}]$ from its closest neighbor (measured from the objects' centers).
The smaller this distance range is, the more regular the objects' arrangement. 

\subsection{Design Decisions}

We made several design decisions when designing our experiment, based on extensive pilot testing and the above theoretical framework.

\subsubsection{Choice of Animation}

Because we primarily study the impact of dynamic changes on perceptual grouping, our main focus when testing dynamic variables lies in what happens during the animated transition itself, and nothing more.
We want to prevent any bias that may be caused by the exposure to the first frame (i.e.\ the initial static state) or the end frame (i.e.\ the final static state).
Hence, for the tasks involving dynamic changes, we prompt the participants with an animation that loops continuously, with a short interruption---a white screen---between two loops.
In other words, the visual objects are never static.

\subsubsection{Choice of Object Number}

We used a restricted classification task~\cite{Garner1974, Ware2012}, which consists of mapping conflicting visual variables to different subsets within the set of visual objects, and then asking participants which grouping is strongest. 
While three objects is more typical when measuring separability~\cite{Maddox1992}, we chose to use four objects grouped by two visual variables in our task because we wanted there to be two clear groups in each trial, as described in Garner~\cite{Garner1974}.
Four is the maximum number of objects that conforms to research in subitizing and object tracking, which states that most people can track up to four moving objects with near-perfect accuracy~\cite{Cavanagh2005}.
In practice, given objects $A$, $B$, $C$, and $D$, we assign the pairs $(A, B)$ vs.\ $(C, D)$ to be similar in variable $V$, and the orthogonal pairs $(A, C)$ vs.\ $(B, D)$ to be similar in $V'$.

Four objects organized in groups of two yields three possible grouping choices.
For simplicity, we give all three groupings as multiple-choice options in a trial.
Two groupings relate to the two respective visual variables ($V$ and $V'$).
The third choice, however, has no meaning, and indicates that a participant perceived the strongest grouping from a non-existent similarity.
If this meaningless choice is favored by participants, this may mean there was a confound, or that none of the visual variables involved cause perceptual grouping.
This is useful for a crowdsourced study such as ours, where we have less control over participants and experimental settings than for a laboratory study.

\subsubsection{Grouping Strength}

For a trial involving two variables $V$ and $V'$, we define the \emph{grouping strength} of $V$ and $V'$ as binary scores---preferred vs.\ not preferred.
If participant indicates that one of the variables is preferred over the other, this results in a grouping strength of $1$ for the former and $0$ for the latter; if neither variable is selected, both are set to $0$.

%% file: 05-user-study.tex
%!TEX root = structanim.tex
%% --------------------------------------------------------------------- 
%% User Study (Methods)
%% ---------------------------------------------------------------------

\section{Method}
\label{sec:user-study}

\begin{figure}[tbh]
  \centering
  \framebox{\includegraphics[width=0.95\columnwidth]{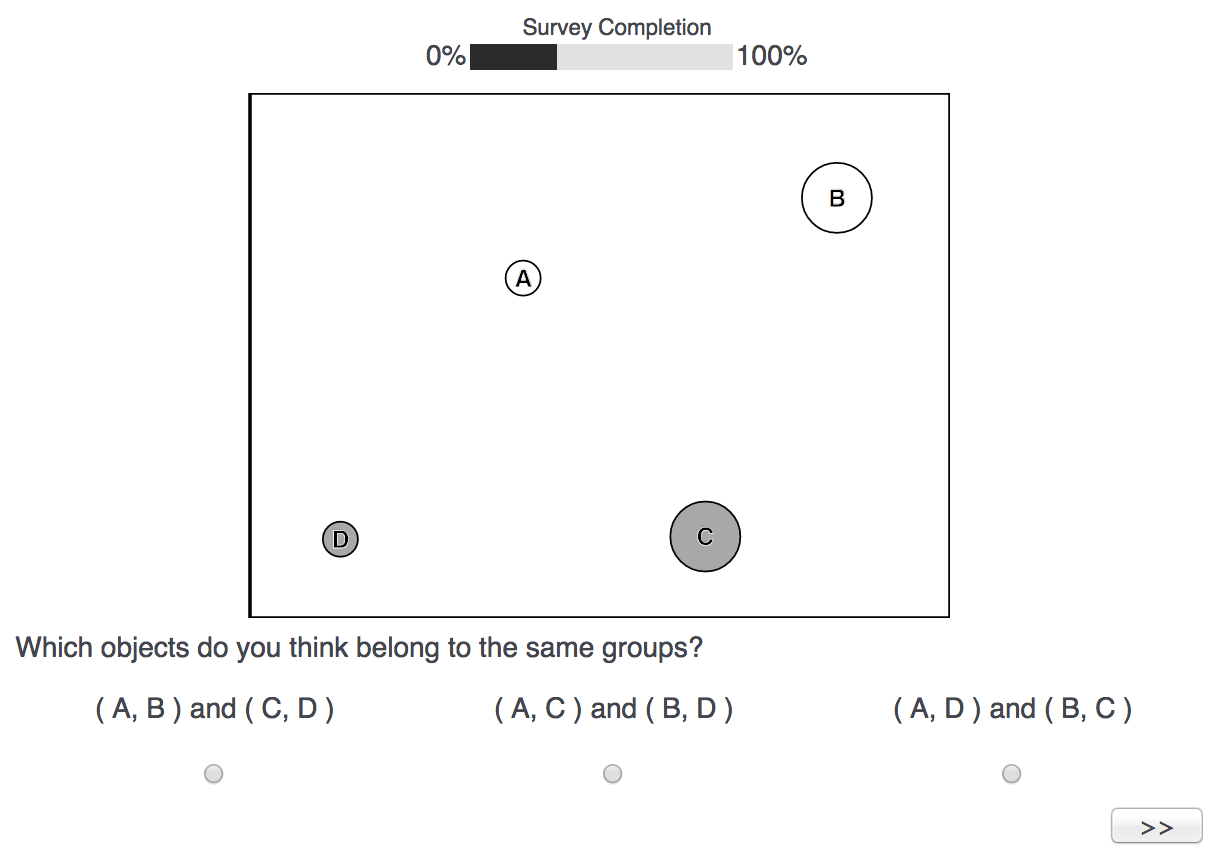}}
  \vspace*{-0.5em}
  \caption{User interface of our study, during a \DS-\SL~task.
    Here, A-B and C-D are pairs of \SL-similar objects, whereas A-D and B-C are pairs of \DS-similar objects, forming the two concurrent possible groupings.}
    \vspace*{-1em}
  \label{fig:ui-exp}
\end{figure}

Based on the above background, we designed a crowdsourced study to investigate our research questions (Figure~\ref{fig:ui-exp}).
The survey can be found at {\small{\surveyURL}} and supplementary materials for all studies at {\small\url{https://osf.io/3zkhv}}.
Here we describe how we generated the tasks; see supplementary materials for further details.

\subsection{Task Generation}

We generate visual objects as circles with a default radius of 10px (relative to the default viewport size of 300$\times$400 pixels), a default luminance of $0.66$ ($\mathrm{luminance}\in 
[0,1]$) and outlined with a black stroke.

To guarantee that the groups are unambiguous, we test the most favorable conditions for each visual variable by adopting a strategy that aims to maximize the intra-group similarity ($\tau_V$), while ensuring a large enough inter-group distance ($\theta_V$) (Table \ref{tab:manipulations}).
To maximize divergence in behaviour while reducing predictability, the parameters vary slightly across conditions.
\SP constraints also vary depending on the maximum size of the visual objects in each condition.
In all cases, $\theta_V$ is well above just-noticeable difference (JND) thresholds, and we believe any systemic impact on the outcome due to these differences is minimal.
For \DP, clusters have randomly generated perpendicular velocities with the same speed, and we enforce minimum pairwise distances between initial and final \SP of all four objects.
For \DL and \DS, opposing groups start at an identical neutral value and then diverge in luminance or size.
We also prevent objects from overlapping or leaving the screen. Depending on the task, the resulting visualization is either static (i.e.\ \SP-\SS, \SP-\SL, \SS-\SL), or a two-second infinitely-looping animation (i.e.\ all other tasks, since they involve a dynamic variable).
We used linear interpolation for \DP, \DS and \DL. 

\begin{table}[t]
    \centering
    \caption{Inter-group\protect\footnotemark \ Distance ($\theta_V$), Intra-group Similarity ($\tau_V$), and Grouping Parameters for the Study. ($\perp$ denotes perpendicular trajectories, $\nearrow$: value increases, $\searrow$: value decreases)}
    \vspace{-5pt}
    \resizebox{0.85\linewidth}{!}{%
    \begin{tabular}{rp{7cm}}
        
        \toprule
        \textbf{Comp} &  \textbf{Parameters} \\%[0.15cm]
         \midrule
        \SS-\DP & $\mathbf{\theta_\mathrm{SS}=10px}$ 
        \small{($\mathrm{SS}_{1}=10px$, $\mathrm{SS}_{2}=20px$)} \newline 
        \normalsize $\mathrm{DP}_1 \perp \mathrm{DP}_2$ 
        \small{($\theta_{\mathrm{SP}_{initial, final}} > 118px$, $53 px/sec < |\mathrm{DP}| < 106 px/sec$)}\\[0.15cm]
        
        \SS-\DL & $\mathbf{\theta_\mathrm{SS}=10px}$ \small
        ($\mathrm{SS}_{1}=10px$, $\mathrm{SS}_{2}=20px$) \newline
        \normalsize $\mathrm{DL}_1 \searrow \mathrm{DL}_2 \nearrow$
        \small($\mathrm{DL}_{1}=0.5\rightarrow0.0$ , $\mathrm{DL}_{2}=0.5\rightarrow1.0$)\\[0.15cm]
        
        \SL-\DP & $\mathbf{\theta_\mathrm{SL}=0.66}$
        \small($\mathrm{SL}_{1}=0$, $\mathrm{SL}_{2}=0.66$) \newline
        \normalsize $\mathrm{DP}_1 \perp \mathrm{DP}_2$
        \small {($\theta_{\mathrm{SP}_{initial, final}} > 57  px$, $53 px/sec < |\mathrm{DP}| < 177 px/sec$)}\\[0.15cm]
        
        \SL-\SP & $\mathbf{\theta_\mathrm{SL}=0.67}$
        \small ($\mathrm{SL}_{1}=0.33$, $\mathrm{SL}_{2}=1.00$) \newline
        \normalsize $\mathbf{\tau_{\mathrm{SP}} < 57  px}$\\[0.15cm]
        
        \SL-\DS & $\mathbf{\theta_\mathrm{SL}=1.00} $
        \small($\mathrm{SL}_{1}=0.00$, $\mathrm{SL}_{2}=1.00$) \newline
        \normalsize $\mathrm{DS}_1 \searrow \mathrm{DS}2 \nearrow$
        \small{($\mathrm{DS}_{1}= 20 px\rightarrow10 px$ , $\mathrm{DS}_{2}=20 px \rightarrow 30 px$)}\\[0.15cm]

        \SL-\SS & $\mathbf{\theta_\mathrm{SL}=0.34}$ 
        \small $(\mathrm{SL}_{1}=0.66$, $\mathrm{SL}_{2}=1.00)$ \newline
        \normalsize $\mathbf{\theta_\mathrm{SS}=10px}$ 
        \small$(\mathrm{SS}_{1}=10px$, $\mathrm{SS}_{2}=20px)$\\[0.15cm]

        \DL-\DS &  $\mathrm{DL}_1 \searrow \mathrm{DL}_2 \nearrow$
        \small $(\mathrm{DL}_{1}=0.5\rightarrow0.0$ , $\mathrm{DL}_{2}=0.5\rightarrow1.0)$ \newline
        \normalsize $\mathrm{DS}_1 \searrow \mathrm{DS}_2 \nearrow$
        \small $(\mathrm{DS}_{1}= 20 px\rightarrow10 px$ , $\mathrm{DS}_{2}=20 px \rightarrow 30 px)$\\[0.15cm]
        
        \SP-\DS & $\mathbf{\tau_{\mathrm{SP}} < 127  px}$ \newline
        \normalsize $\mathrm{DS}_1 \nearrow \mathrm{DS}_2 \searrow$
        \small $(\mathrm{DS}_{1}= 20 px\rightarrow10 px$, $\mathrm{DS}_{2}=20 px \rightarrow 30 px)$\\[0.15cm]
      
        \SP-\DL & $\mathbf{\tau_{\mathrm{SP}} < 57  px}$ \newline
        \normalsize $\mathrm{DL}_1 \searrow \mathrm{DL}_2 \nearrow$
        \small $(\mathrm{DL}_{1}=0.5\rightarrow0.0$ , $\mathrm{DL}_{2}=0.5\rightarrow1.0)$\\[0.15cm]
        
        \SP-\SS &  $\mathbf{\tau_{\mathrm{SP}} < 85  px}$ \newline 
        $\mathbf{\theta_\mathrm{SS}=15px}$
        \small $(\mathrm{SS}_{1}=10px$, $\mathrm{SS}_{2}=25px)$\\[0.15cm]
        
        \DP-\DS & 
        \normalsize $\mathrm{DP}_1 \perp \mathrm{DP}_2$
        \small $(\theta_{\mathrm{SP}_{initial, final}} > 113  px$, $53 px/sec<|\mathrm{DP}| < 152 px/sec)$ \newline
        \normalsize $\mathrm{DS}_1 \searrow \mathrm{DS}2 \nearrow$
        \small$(\mathrm{DS}_{1}= 20 px\rightarrow10 px$ , $\mathrm{DS}_{2}=20 px \rightarrow 30 px)$\\[0.15cm]
        
        \DP-\DL & 
        \normalsize $\mathrm{DP}_1 \perp \mathrm{DP}_2$
        \small ($\theta_{\mathrm{SP}_{initial, final}} > 57  px$, $53 px/sec < |\mathrm{DP}| < 106 px/sec$) \newline
        \normalsize $\mathrm{DL}_1 \searrow \mathrm{DL}_2 \nearrow$
        \small{($\mathrm{DL}_{1}=0.5\rightarrow0.0$ , $\mathrm{DL}_{2}=0.5\rightarrow1.0$)}\\[0.15cm]
        
        \bottomrule
        
    \end{tabular}
    }
    \vspace{-5pt}
    \label{tab:manipulations}
\end{table}

\footnotetext{Except initial and final \SP distance constraints in \DP conditions, which are pairwise between all four objects}

\subsection{Attention Trials}

Although crowdsourced graphical perception experiments are a powerful tool~\cite{Heer2010}, special care is required for large-scale studies~\cite{Kittur2008}.
A typical approach is to add attention trials---i.e.\ trials where an obviously correct answer exists---to filter out careless responses.

We designed attention trials as trials similar to the regular ones where, instead of opposing two visual variables against each other, we make them work together.
This results in a single meaningful grouping defined by both visual variables, and participants who chose other grouping options should be regarded as potentially insincere.

Three types of attention trials were implemented using the following combinations: {\SS-\DP}, {\SP-\DL}, and {\SL-\DS}, each of which was repeated twice.
It should be noted that these are virtually inseparable from regular trials, so workers are unlikely to be able to game the study, which was confirmed by some of the participants' comments.

\subsection{Procedure}

Participants were first asked to give informed consent, and were given a calibration guide: we asked them to maximize the browser window and zoom to make the experiment interface visible and avoid scrolling.

Participants were instructed that their task consisted of selecting the pairwise grouping of these objects that best fit their intuition among three proposed groupings---presented as a list of radio buttons.
Participants were forced to pick one grouping before moving to the next trial.
They could change their mind until pressing the ``Next'' button, and were instructed to disregard completion time, and to  take breaks. 

Participants were explicitly instructed to ``not spend too much time thinking since there is no correct or wrong answer'', and were strongly encouraged to stick to their first impression.
Finally, they were warned that they would be given attention tests, and that we may reject their responses based on these tests.
After being introduced to the task, participants were asked to practice on a set of three pre-defined trials.

Participants were then asked to perform the experiment.
Depending on whether the task involved the manipulation of a dynamic variable or not, participants were shown either a static visualization, or a two-second animated transition that played automatically and looped infinitely until a grouping answer was submitted.
Participants had no control over animation.
No feedback was given on submitted answers.

After finishing all trials, participants filled out a demographic survey and were given the opportunity to provide freeform comments.

\subsection{Participants}

We posted the experiment on \href{https://www.mturk.com/mturk/welcome}{Amazon Mechanical Turk}.
100 participants (46 female), all  \href{https://www.mturk.com/mturk/help?helpPage=worker#what_is_master_worker}{Turk Masters}, responded, and were paid $\$2$.

\subsection{Apparatus and Setup}
    
The experimental survey was created on the \href{http://www.qualtrics.com}{Qualtrics} platform.
Custom JavaScript code built using \href{http://d3js.org}{D3} was inserted to generate animations.
Since the study was crowdsourced, we did not have control over the study apparatus, but we logged information about the screen resolution and the web browser used during the study.
All participants used Google Chrome or Mozilla Firefox browsers as advised, at a screen resolution ranging from $320\times480$ (iPhone) to $1920\times1200$ pixels.

\subsection{Experimental Design}

Our independent variables were the visual variables \DP, \DS, \DL, \SP, \SS and \SL~combined into pairs to form the 12 tasks described above.
Each type of task was repeated three times for each participant.
Our experiment factors were as follows:

\vspace*{-0.5em}
\begin{table}[!ht]
\centering
  \setlength{\tabcolsep}{1pt}
  \begin{tabular}{r r l}
    & 100 &~ crowdsourced participants\\
    $\times$ & 12 &~ {\sc Tasks} (pairs of \DP, \DS, \DL, \SP, \SS, \SL)\\
    $\times$ & 3 &~ repetitions  \\
    \hline
    \multicolumn{2}{c} 3,600 & trials (excluding practice trials)
  \end{tabular}
\end{table}
\vspace*{-0.5em}

Trials were grouped into blocks of six, and each block was followed by one attention trial.
The order of trials, regular and attention ones, was randomized across participants.
For each trial, we recorded the visual variables $V$ and $V'$, their grouping strength (i.e.\ the binary score of whether the variable was preferred), and the completion time.

\subsection{Hypotheses}

\begin{itemize}[itemsep=0pt, topsep=0pt]
\item[H1]\textit{The LCF extends to dynamic luminance and dynamic size.}
This prediction follows from our discussion above, from Wertheimer himself~\cite{Wertheimer1923} and prior studies on correlated modulations of luminance as a figure-ground segmentation factor~\cite{Sekuler2001, Wagemans2012, Ware2005}.
  
\item[H2]\textit{The grouping strength of Common Fate is higher than Proximity.}
Animation has high urgency~\cite{Tversky2002}, so we think it will outperform even the Law of Proximity.
  
\end{itemize}

%NOTE: Section 6 starts at the top of page 6

%% file: 06-results.tex
%% --------------------------------------------------------------------- 
%% Results
%% ---------------------------------------------------------------------
\section{Results}

Drawing on current best practices in HCI~\cite{Cumming2013, Dragicevic2014}, our data analysis is based on estimation, i.e.\ effect sizes with confidence intervals~\cite{Cumming2005}.
This approach also aligns with the latest recommendations from the APA~\cite{APA}, and has been successfully applied in prior work evaluating animated transitions~\cite{Chevalier2014}.\footnote{Dragicevic~\cite{dragicevic2016fair} provides more details on fair statistical practices for HCI.}

\subsection{Data Verification}

Before running analyses, we performed a series of verifications to filter out invalid data and careless responses.
We counted the number of failed attention trials.
Out of 6 trials, only 8 participants failed 1 or 2 of such trials. Interestingly, several participants commented explicitly on the attention trials, \eg \quote{Despite watching with a paranoid intensity I never noticed any of the attention checks.}

We also looked at answers that did not correspond to either visual variable involved in the trial, which can be seen as incoherent groupings.
This data helps us verify if participants understood the instructions correctly.
Twenty-one participants picked an incoherent choice once (out of the 42 trials), 5 picked an incoherent choice twice, and a single participant picked such an option 9 times in total.
Given a total of 3,600 trials, this gives us confidence that our results are sound.

\subsection{Dynamic vs.\ Static Groupings}
\label{sec:common-DynVsStatic}

\begin{figure}[htb]
    \centering
    \includegraphics[width=0.4\columnwidth]{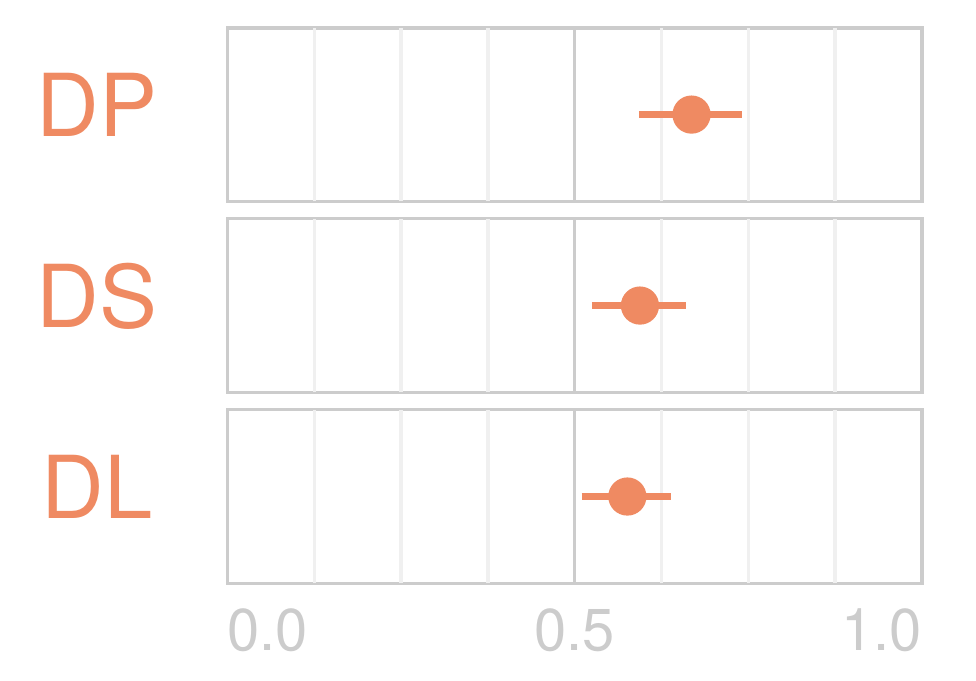}
    \vspace*{-0.6em}
    \caption{Mean grouping strength for the dynamic variables (\DP, \DS, \DL) in the context of a conflicting static grouping.
    Error bars are $95\%$ CIs.
    That all CIs fall to the right of the halfway mark (0.5) shows that these dynamic variables have stronger grouping strength than static ones.}
    \vspace*{-1em}
    \label{fig:RQ1}
\end{figure}

In our study, we first set out to determine whether the LCF extends to dynamic luminance or size.
Figure~\ref{fig:RQ1} shows the overall grouping strength of all three \emph{dynamic} variables (one per row) when competing against \emph{static} variables only (see Figure~\ref{fig:comparison} for per-task results).
The grouping strength of each dynamic visual variable was assessed by aggregating all trials where that variable was in competition with a static visual variable (i.e.\ 2 tasks $\times$ 3 repetitions $ = $ 6 data points per variable per participant), yielding a total of 100 data points for each of the \DP, \DS, and \DL~variables.
Point estimates and $95\%$ confidence intervals were computed using bootstrapping~\cite{Kirby2013}.
In the figures, higher values (to the right) mean higher grouping strength.
Overall, the grouping strength of a variable can be understood as the probability that the variable will be most influencing perceptual grouping over another, conflicting visual grouping cues.
Hence, a grouping strength above $0.5$ in Figure~\ref{fig:RQ1} indicates a higher influence of the dynamic variable on grouping over that of the tested static variable.

Overall, our data confirms the LCF for motion (\DP).
Furthermore, the data leaves no doubt as to the strong positive effect of \DL~and \DS~on perceptual grouping for the type of tasks we tested, since all point estimates and associated CIs fall in the right half of the plot (i.e.\ above $0.5$).
In other words, grouping based on the dynamic behavior was chosen in more cases than cases it was not chosen.

\subsection{Relations Between Gestalt Laws}
\label{sec:common-btwnGestaltLaws}

\begin{figure*}[htb]
    \centering
    \includegraphics[width=\textwidth]{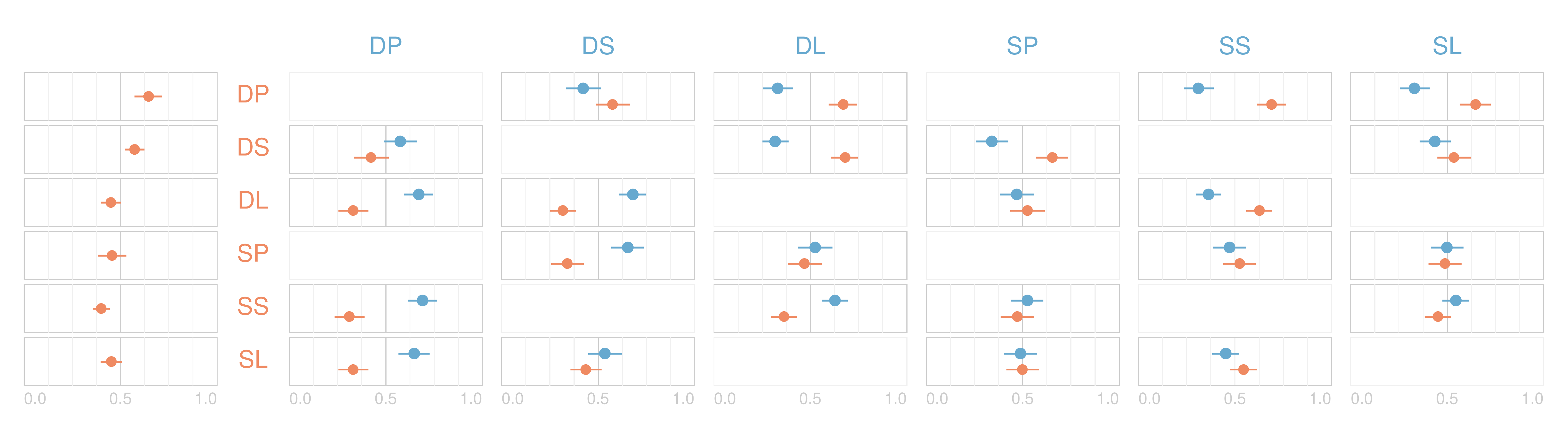}
    \vspace*{-2em}
    \caption{Summary of the mean grouping strength for all visual variables examined in our main experiment.
    Error bars are $95\%$ CIs. 
    The left column shows the mean grouping strength per variable, for all comparisons it was involved in.
    The fact that the CIs of \DP and \DS fall to the right of the plots shows that they have stronger grouping strength than any other tested variables.
    Each cell in the rest of the table shows the mean grouping strength of the corresponding visual variable in the row (orange) vs.\ that of the visual variable in the column (blue), for that task only.}    
    \label{fig:comparison}
    \vspace*{-0.5em}
\end{figure*}

Our second research question concerned the relationship between different Gestalt laws.
Figure~\ref{fig:comparison} shows a detailed summary of the results of our experiment.
The leftmost column shows the overall grouping strength of each visual variable.
The grouping strength was assessed by aggregating all trials where the corresponding visual variable was involved (i.e.\ 4 tasks $\times$ 3 repetitions $ = $ 12 trials per variable per participant) across participants, yielding a total of 100 data points for each visual variable.
A higher value means a stronger grouping power. 

While it is clear from the plots (left column) that \DP~and \DS~have stronger grouping power overall (CIs $>$ $0.5$) than \DL, \SS, and \SL~(CIs $<$ $0.5$), no strict order can be established between all 6 visual variables.

In the right part of Figure~\ref{fig:comparison} (columns 2-7), a tabular view similar to Table~\ref{tab:tasks} summarizes all comparison tasks (one per cell).
Each cell corresponds to the comparison of one visual variable (one per row) against another visual variable (one per column).
The effect of each visual variable on grouping was assessed by performing a contrast, i.e.\ aggregating all trials corresponding to that comparative task for each of the two visual variables involved.

This detailed view allows us to refine our comparative assessment.
Reviewing all visual variables (i.e.\ reading the table column-by-column), we can make all possible pairwise comparisons.
Considering grouping by dynamic behavior, \DP~has a clear observable effect on perceptual grouping when in competition with \DL, \SS, or \SL.
\DS~also exhibits a strong effect when competing against \DL~or \SP.
\DL~has stronger grouping power than \SS.
As for the static variables, \SP~ appears weaker compared to \DS;
it is, however, a fair competitor to \DL, as well as \SS~and \SL.
\SS~is clearly not as strong a grouping factor as variables \DP~and \DL, but is as strong of a factor as \SP and \SL.
Finally, \SL~has a weaker effect on grouping than \DP; however, it has similar grouping power to \DL, \SP, and \SS.

\subsection{Participant Feedback}

Reviewing the open comments that participants entered at the end of study revealed that the grouping difficulty diverged between participants.
Some found the task easy to do, while many others felt it was hard to decide on a grouping.
Reviewing comments of the participants who struggled reach a decision, we discern that pairing was challenging mainly because they found different possible ways to group objects: \quote{It was sometimes hard to determine groups because you could associate them together in a few different ways.}
The conflicting---both valid---options made them hesitant as to what visual factor would yield the best group: \quote{It was hard to decide which characteristic made it the best pairings,} and \quote{It was not always easy to decide.
I tried to prioritize...}
These comments reinforce the idea that perceptual grouping of several groups---where none of the visual variable dominates the other---is possible in some cases.
The forced choice imposed in our experiment may have conflicted with participants' perceptions.

Some participants went beyond giving general comments and explained the grouping strategy they adopted during trials.
Several comments confirmed traditional LCF, e.g., \quote{I tended to group those together that were moving in the same direction,} and \quote{Anytime the motion was the same I pretty much felt that was the highest criteria for what was a group.}
Some manifested a more generalized definition of common fate: \quote{I would look for ones that were doing the same thing---such as changing size---and group those.}
Overall, as suggested by our quantitative results, no strict order between the variables emerged.

%% file: 07-followup.tex
%% ---------------------------------------------------------------------
%% Followup Study
%% ---------------------------------------------------------------------
\section{Follow-up: Common Fate for Trend Visualization}
\label{sec:followup}

The main study in this paper was designed to maximize internal validity by carefully controlling and balancing each experimental factor, at the price of compromising ecological validity.
We focused on a small number of points, and, in order to guarantee observable effects, we enforced the best possible conditions for each visual variable to create distinct groups.
Other possible scenarios include a wider range of values as initial and final states (\eg, shades of gray instead of white or black), making the perceptual identification of groups more difficult.

To provide complementary findings applicable to more realistic data visualization settings with a larger number of visual objects, we also report on a follow-up study assessing the LCF for two types of trend visualizations: an animated scatterplot inspired by Gapminder,\footnote{\url{http://www.gapminder.org/}} as well as a dynamic thematic map of election data.

The focus of this study was on obtaining nuanced qualitative insight into more complicated use cases of the notion of common fate, to further our understanding of what influences grouping in a dynamic visualization.
Thus, we limit this follow-up experiment to the dynamic variants of the three visual variables: \DP, \DS, and \DL. 

In both visualizations, the magnitude of the change in the dynamic variables was mapped to the change in the underlying data variables.
Color scales used linear interpolation in the L*a*b* color space.
Animated transitions lasted 2 seconds and used linear interpolation.
Participants could replay the animation using a 'replay' button.
They entered written explanations for each question, but were also encouraged to think aloud, and to elaborate verbally on their written answers. 

\begin{figure*}[htb]
  \centering
  \includegraphics[width=0.98\linewidth]{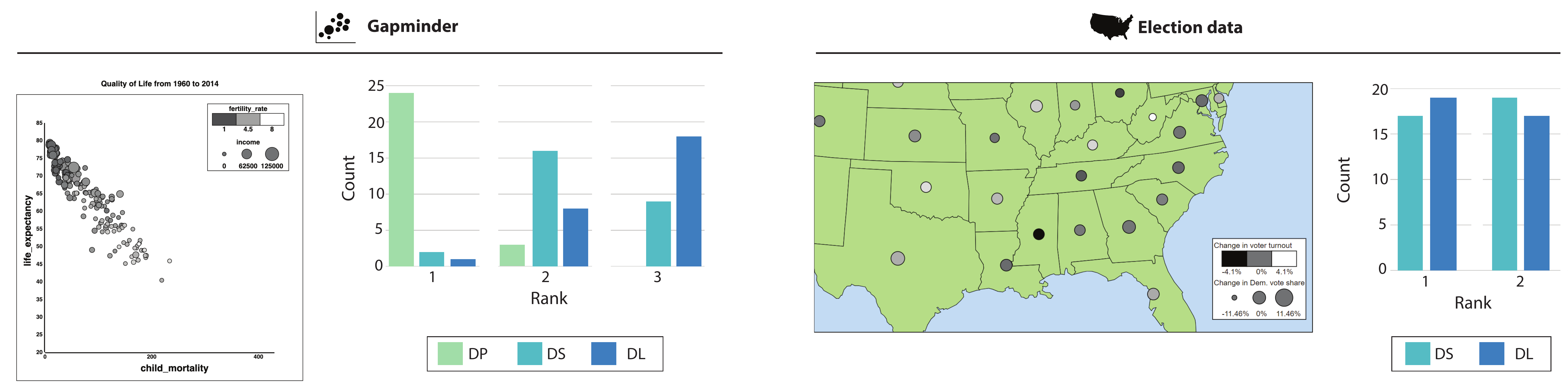}
  \vspace{-4mm}
  \caption{Grouping strength ranking results (numerical answers to Q3$^5$) for the three dynamic variables (\DP, \DS, and \DL) in the context of a Gapminder-like animated scatterplot visualization (left) and an animated thematic map of U.S.\ Presidential election data (right).}
  \label{fig:follow-up-3}
  \vspace*{-1.1em}
\end{figure*}

\subsection{Gapminder-like Animated Scatterplot}

In the animated scatterplot, each bubble represents a country and encodes four data dimensions: life expectancy, fertility rate, child mortality, and income (Figure~\ref{fig:follow-up-3}, left)~\cite{GapminderData:19}. 

In this condition, we presented participants with three different configurations of the scatterplot that we generated by varying the visual encoding of the four data dimensions.
Life expectancy was always assigned to vertical position.
The other variables were each assigned to horizontal position, size, and luminance exactly once.
The experimental interface is available at: {\small\url{http://tiny.cc/gapminder_plot}}.

During each animated transition, all three visual variables change simultaneously to portray the change in the data between 1960 and 2014 (linearly interpolating between the two years).
The visualization viewport was 600$\times$600 pixels, the radii of points varied between $[4px, 14px]$, and luminance varied between $[0.33, 1.00]$ since occlusion made distinguishing countries difficult when using pure black.

\subsection{Election Data Bubble Map}

In the map visualization (Figure~\ref{fig:follow-up-3}, right), an animated bubble is displayed on top of each state to represent two data dimensions: change in voter turnout~\cite{VoterTurnout12, VoterTurnout16} and Democratic vote share from 2012 to 2016~\cite{VoteShare12, VoteShare16}. 
All points start out with an identical color and radius to minimize the influence of the initial state. 
 
The study involved four different configurations of the map visualization that we generated by varying both the visual encoding of the two data dimensions and the section of the map that was displayed. Every participant saw the same four map sections, which each showed roughly one quarter of the U.S.\ (with some overlap) to reduce training effects between trials.
The two data dimensions were mapped to one of \DL or \DP in each trial.
Hence, the four configurations presented the combination of two different instances of visual encoding and four different map sections.
The viewport size was 800$\times$400 pixels, the radii of points varied between $[4px, 14px]$, and luminance varied between $[0.0, 1.0]$.
The experimental interface can be accessed at: {\small\url{http://tiny.cc/election_map}}.

\subsection{Method}

We recruited 10 participants ($M_{age} = 24$, 4 female) using in-course announcements and university message boards.
Compensation was \$10 or course credit, and the experiment lasted 45--60 minutes.

The order of the two visualization conditions was counterbalanced between participants.
Participants answered all questions for every trial in each of the two conditions (total number of questions was 21):

\begin{enumerate}[nosep]
    \item[Q1.] What patterns do you see in the visualization?
    Describe (at least) 2--3 groups of points that behave in the same way.

    \item[Q2.] Which [country/state] is closest in behavior to the selected one?
    
    \item[Q3.] Which dynamic variable(s) create the most visible patterns?
\end{enumerate}

The experiment was audio recorded.
The participants were instructed to give written explanations for each of their answers, and were encouraged to give verbal details on their analytic process. 

\subsection{Results}

Results were based on a holistic analysis of participants' input, including written answers to the experimental trials, transcriptions of think-aloud comments, and numerical answers to survey questions.
The author who conducted the experiment developed a list of top-down themes that were used to qualitatively analyze the data.

Overall, all participants successfully identified patterns using all three dynamic variables in response to Q1; \eg, \quote{small group of points moving horizontally [...] while also getting darker, but no significant change in size}  (P8), and \quote{one set of points increase in size, another (possible overlapping) set go black} (P6). 

In their answers participants also showed that they could effectively map dynamic changes in the visual variables to meaningful trends in the underlying data; \eg, \quote{points on the right got darker, while on average also increasing in size slightly.
This indicates that there was a negative change in [Democratic] vote share, with a small increase in voter turnout.} (P8), and \quote{Life expectancy for richer people gets higher} (P7).
Overall, participants tended to reason in the data space for the pattern identification task (Q1), while they combined both data space and visual space reasoning for the closest behavior task (Q2).

Most of the participants successfully performed conjunction search in response to Q2, by finding data points that changed similarly in all dynamic variables simultaneously.
However, others used only one or two visual variables in the Gapminder example: \eg, P3 used only \DP and P5 used only two variables at a time (\DS and either \DL or \DP).

Participants made semantically meaningful choices using the three dynamic variables with no access to information about the identity of the countries (\eg, Qatar was most often identified as closest in behavior to Kuwait while Thailand and Brazil were most commonly identified as similar to Vietnam).
Several participants described focusing on different variables sequentially in this task, e.g., \quote{I first look at the change in size [...] and then I look at the brightness} (P4).

\subsubsection{Relative Strength of Dynamic Variables}

Numerical answers to Q3\footnote{P5's answers for the map condition were reversed because her written justification made clear that she had inverted the scale.
P4's answers were excluded because he did not assign unique ranks to the dynamic variables.} (Figure~\ref{fig:follow-up-3}) provide a ranking that can be compared with the results of our crowdsourced study. 

In the Gapminder condition, \DP was overwhelmingly ranked as the strongest variable.
\DS appeared to be the next strongest, and was ranked first or second twice as often as \DL.
This deviates slightly from the first study, where \DP was only marginally stronger than \DS.

In the map condition, however, while \DL was ranked as the strongest variable slightly more often than \DS, there was no clear overall trend.
Participants commented that it was difficult to rank these variables, \eg, \quote{Between color and size, both seem equally prominent} (P5).
This again differs from past results, where \DS dominated \DL. 

Differences in the underlying data caused rankings to vary between trials for the same participant, with only two participants assigning the same ranks to \DP and \DS for all four maps, and only one participant assigning completely consistent ranks in the Gapminder condition.

\subsubsection{Dynamic Behavior}

Participants used both the dynamic behavior of the points and their initial and final states when identifying trends; \eg, P7 felt two countries were close in behavior \quote{because position-wise they're converging to the same place and [...] color-wise they're also doing [the] same.} 
P9 focused on dynamic behavior; i.e., \quote{they didn't start [...] similar to each other, but their rapid growth was similar.}
When describing similarities in dynamic behavior, participants commented mostly on direction and magnitude of the change: \quote{It's moving the same way, moving in the same direction at nearly the same speed} (P6).
The perceived synchrony of change also seemed important for dynamic behavior: \quote{they changed almost at the same time [...] the brightnesses are getting darker [in the] same direction at the same time} (P5).

\subsubsection{Proximity}

Proximity was mentioned several times as an important cue for group formation.
Though our definition of common fate similarity in \S~\hyperref[sec:similarity]{\ref{sec:similarity}} is independent of proximity (\SP-similarity), participants reported that it was easier to group data points together using one of the dynamic variables if they were close together; \eg, \quote{They are also close to each other, so it was easy to detect them} (P2), \quote{I guess I have greater confidence too, when it's closer} (P10).
In addition, when answering Q2, P4 reported focusing on countries near the target in the Gapminder condition, and P5 used proximity to the target to break ties between otherwise similar candidates in the map condition.
While the first experiment showed that \DP and \DS have stronger mean grouping strengths than \SP, these observations suggest that \SP-similarity could play a mediating role in the perception of all three dynamic variables.

While close distances seemed to facilitate comparison and group identification for some participants, sparsity also played an important role in detecting dynamic behaviors: \quote{This one is more noticeable than other circles because it's isolated} (P2); \quote{Position is very obvious.
The circles are becoming extremely sparse!} (P7).

\subsubsection{Cardinality}

Several participants noted that the number of points that underwent a certain dynamic change increased the visibility of the patterns variable produced.
\quote{Brightness was [...] strongest because there were three points that changed in brightness dramatically while only one point changed in size} (P8), and \quote{The size was the strongest due to the larger amount of circles in this cluster} (P9).
On the other hand, a few participants felt that a small number of outliers that changed significantly in one dynamic variable could influence them to find the rest of the changes in that variable greater as well: \quote{those circles that move quickly, that kind of biases my mind} (P2), and \quote{The white circle on the upper right catches my attention more than any other circles, so I tend to think the size is more noticeable} (P2).
Still, participants were also able to detect patterns involving a small number of data points: \quote{Only a few circles both change income and life expectancy} (P2), and \quote{there is a group of 2 or 3 points which started in the bottom left and moved vertically upwards while slightly decreasing in size} (P8).

\subsubsection{Interactions Between Dynamic Variables}

One participant felt that because \DP drew his attention, he was more likely to notice changes in \DS in the vicinity: \quote{my eyes keep following that trend [...] the trend of position which correlates to stronger size changes} (P10).
\DP was also sometimes perceived as interfering with the other two variables: \quote{the color and the size [are] so hard to see when there's so much stuff and it's moving so fast} (P1). 

%% file: 08-disc.tex
%% --------------------------------------------------------------------- 
%% Discussion
%% ---------------------------------------------------------------------
\section{Discussion}

Our work focuses on exploring the meaning of the seemingly straightforward concept of ``common fate'' in Gestalt psychology.
Is that term merely shorthand for ``visual elements that move in the same way,'' or does it suggest a more fundamental meaning: visual elements behaving in the same way, regardless of what that behavior is? 
Put differently, we were interested in learning which dynamic behaviors, beyond common velocity, would create the perception of grouping in a dynamic visualization.
Our secondary objective was to determine how these dynamic groupings interact with static visual properties such as proximity, luminance, and size of graphical objects.
The results from our crowdsourced study give a quantitative basis to begin answering these questions, while our follow-up study offers insight into how these results might generalize to real-world settings.

Overall, our results confirm H1, that the Law of Common Fate applies to dynamic luminance and dynamic size, and that these variables can be used for analyzing trends in realistic dynamic visualization scenarios.
We further show that users can effectively perform conjunctive queries using two or three dynamic variables simultaneously.

We confirm, as shown in Figure~\ref{fig:RQ1}, that the three dynamic variables (\DP, \DL, and \DS) present a clear overall benefit compared to static visual variables, and in particular confirm H2 (that the Law of Common Fate is stronger than the Law of Proximity) for \DP and \DS (Figure~\ref{fig:comparison}). 

We also develop some notion of the relative strengths of the three dynamic variables;
in  Figure~\ref{fig:comparison}, \DL is clearly outperformed by \DP and \DS while there is no clear distinction between these latter two.
However, in our second study, we observed some perturbation of the ranking---\DL and \DS, which had been clearly different in grouping strength, became difficult for participants to rank when evaluated in isolation in this study.
Similarly, \DS and \DP were roughly equivalent in the main study, but \DP outperformed \DS in the second one. 

These deviations may both result from \DS becoming weaker under less idealized conditions, where larger number of points each change by smaller magnitudes.
The fact that the trends in the election data set created clusters that were less distinct and discriminable than in the first study may have also led to the decrease in the dominance of the \DS condition.
Realistic design scenarios may also put limits on parameter values that might affect relative grouping strengths.
For example, design constraints forced us to use a smaller dynamic range for luminance in the Gapminder visualization. 

The fact that \DS still dominated \DL in the Gapminder condition may be related to the interference between \DP and the other variables.
In the real world, objects usually grow or shrink in our field of view when moving closer or further~\cite{Nesbitt2002}.
It may well be the case that, when compared to \DL, \DS benefits from our ability to perceive such motion and so is more effective when employed in conjunction with \DP.
This aligns with prior work on motion decomposition~\cite{Friedrich2002, Nesbitt2002}.

\subsection{Implications for Dynamic Information Visualization}

The two studies have several design implications that can guide the use of  common fate groupings in visualization.

\emph{Encoding dynamic groups in visualizations.}
Choosing visual variables to emphasize groups is highly constrained 
by the visual encoding of the visualization.
For example, if \SP is already mapped to an attribute, employing \DP for grouping is not possible, since it may be perceived as a change in this underlying attribute's values.
The fact that \DL and \DS create the impression of common fate thus greatly increases the range of possibilities for encoding dynamic groups.

\emph{Revealing several groups.}
When encoding two variables in conjunction, use variables that are similar in perceptual grouping strength.
For example, in our follow-up experiment, participants found \DS/\DL and \DP/\DS pairings more tractable than the \DP/\DL pairing, which is consistent with the main study results. 
This also implies that \DL would be the most appropriate dynamic variable for use in conjunction with \SP, and \DS the most appropriate for use in conjunction with \SL (a combination that has been previously investigated in graph drawing~\cite{Friedrich2002}).

\emph{Making a single group salient.}
When used to focus attention on a single group, the visual variable with strongest grouping power should be preferred.
For example, if objects have arbitrary size and semantically meaningful static position and dynamic luminance (i.e., \SP  and \DL encode values), a sub-group has most chance to pop out if highlighted using \DS rather than \SS, the equivalent static variable.

\subsection{Future Work}

Some limitations to our studies suggest further investigations into the role of dynamic visual properties in perceptual grouping.

\emph{Interference and separability.}
Robust determination of relative grouping strength relies on the assumption of separability---i.e., that participants can make independent judgments about each variable~\cite{Ashby1994, Garner1974}.
While the static variables are known to be separable or mostly separable~\cite{Ware2012}, the separability of the dynamic variables is less well-characterized.
Our work mostly supports the separability of the tested combinations, especially comments from the closest behavior task, where participants described attending to each dimension sequentially.
However, other comments showed that \DP could be distracting, impeding participants' ability to perceive other variables, which is consistent with prior work on motion silencing~\cite{Suchow2011}.
Since participants repeatedly commented that \DP seemed to draw their attention, it is possible that large changes in position blind participants to smaller changes in color or size, forcing recourse to focused attention to perceive these trends.
Interestingly, prior work in vision science \cite{McKee1986} has found that \DL does not interfere with accurate velocity perception, suggesting that these variables may be \emph{asymmetrically separable}~\cite{Ashby1994}.
The degree of interference between the pairs of variables treated here requires future investigation.
The speeded classification tasks described by Ashby and Maddox~\cite{Ashby1994} would help give a complete account.

\emph{Parameter values.} The main study would have benefited from more principled selection of parameter values (c.f., \cite{Szafir2017}), since these may have affected grouping strengths.
While our crowdsourced study enabled us to recruit a larger number of participants than would have otherwise been feasible, it also adds some variability since we were unable to control resolution or display parameters that are important for a psychophysical study.
Repeating our experiments in a lab scenario with a wider and better controlled range of parameters would strengthen the conclusions.
On the other hand, Heer and Bostock~\cite{Heer2010} have shown that many graphical perception effects from laboratory studies can be replicated in a crowdsourced setting.
Furthermore, relative grouping strengths of separable variables are known to be ``fairly impervious to manipulations of discriminability''~\cite{Garner1974}, giving a degree of confidence that our results would match such a replication.

\begin{figure}[!tb]
    \centering
    \includegraphics[width=0.77\linewidth]{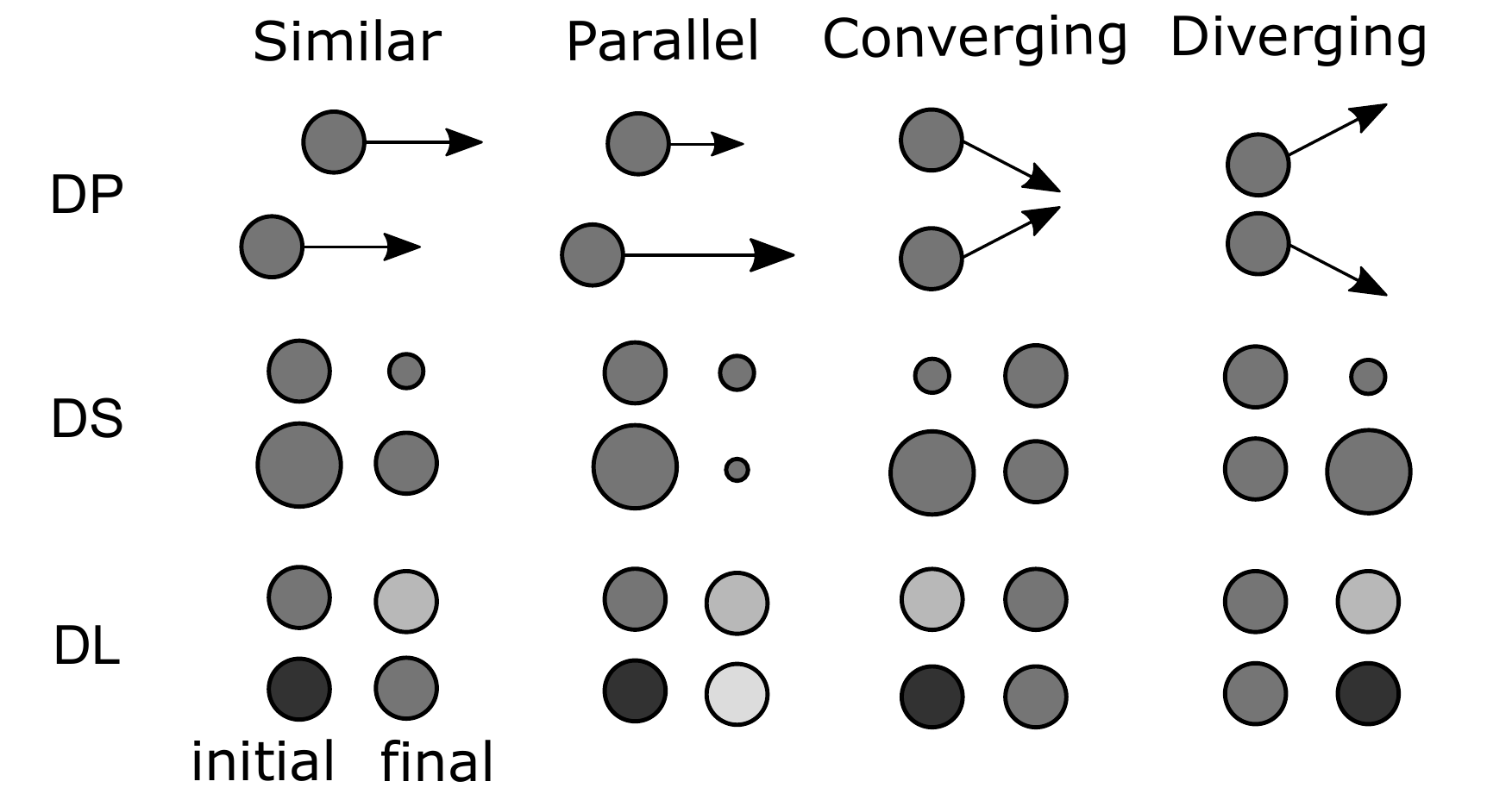}
    \caption{Combinations of dynamic behavior and initial/final conditions that might emerge in realistic visualization scenarios.}
    \label{fig:dyn_beh}
\end{figure}

\emph{Rich interpolations and dynamic relationships.} In this work, we studied monotonous, linear interpolations in ordered space.
More sophisticated transitions, e.g., rhythmic patterns (blinking, pulsing, vibration) or exaggerations (overshoot) may create further groupings that follow LCF.
Moving from the tightly controlled first experiment to a scenario where the magnitude and granularity of transitions were determined by underlying data created more complex dynamic relationships.
In describing their decision process, participants made use of a diverse range of features, including perceived direction, speed, magnitude and synchrony of changes, as well as similarity of initial and final states.
Figure~\ref{fig:dyn_beh} shows a few of the possible emergent combinations of these features. Future work should quantitatively investigate their possible effects on grouping strength.

%% file: 09-conclusion.tex
%% --------------------------------------------------------------------- 
%% Conclusion and Future Work
%% ---------------------------------------------------------------------
\section{Conclusion} 

We have presented results from a crowdsourced perception task involving four visual objects with varying static and dynamic visual properties. By orthogonally pairing these objects, we were able to assign conflicting visual behaviors to two sets of pairs, and determine which of these behaviors resulted in stronger perceived~grouping.

Ultimately, our results confirm empirically that the Law of Common Fate is not restricted to mere motion, and that dynamic luminance and dynamic size also have strong grouping power.
Our follow-up study demonstrated how the relative strength of dynamic visual variables might shift when applied to more realistic visualization scenarios, and showed that participants were capable of making use of these variables for pattern identification and conjunction search.

In the future, we anticipate operationalizing these results to design animated transitions that are easier to comprehend. 
Visual grouping can be used to create structured animations that reduce the number of objects to track~\cite{Cavanagh2005, Pylyshyn1988}.
As such, our results uncover one route forward in improving the effectiveness of animated data graphics.